\documentclass[runningheads]{llncs}
\usepackage[T1]{fontenc}
\usepackage{graphicx}
\usepackage{xcolor}
\usepackage{amsfonts}
\usepackage{amsmath}
\usepackage{lmodern}

\usepackage[latin1]{inputenc}
\usepackage{url}
\usepackage{setspace}
\usepackage{listings}
\usepackage{hyperref}
\usepackage{booktabs}
\usepackage{framed}
\usepackage{pifont}% http://ctan.org/pkg/pifont
\newcommand{\cmark}{\ding{51}}%
\newcommand{\xmark}{\ding{55}}%

\lstnewenvironment{AnBxlisting}[1][]
{\lstset{language=C,basicstyle=\ttfamily\scriptsize,frame=single,
 keywords=[1]{Agent, Number, Function, Symmetric_key, PublicKey},
 keywords=[2]{Protocol, Types, Knowledge, Goals, Actions},
 keywords=[3]{section, inits, rules, step, initial_state, attack_state, not, where, authenticates, on, secret, between},
 keywordstyle=[1]\bfseries\color{violet},
 keywordstyle=[2]\bfseries\color{magenta},
 keywordstyle=[3]\bfseries\color{teal},
 morecomment=[l]{\#}, % line comment
 morecomment=[l]{\%}, % line comment
 literate=%
    {->}{\begingroup{\color{blue}$\rightarrow$}\endgroup}2%
    {\{}{\begingroup{\color{blue}\{}\endgroup}1%
    {\}}{\begingroup{\color{blue}\}}\endgroup}1%
    {\{|}{\begingroup{\color{red}$\{\vert$}\endgroup}2%
    {|\}}{\begingroup{\color{red}$\vert\}$}\endgroup}2%
    {*}{$\bullet$}1,
 commentstyle=\itshape\color{gray},
 identifierstyle=\color{black},
 stringstyle=\color{brown},showstringspaces=false,
 caption={#1}
 } }
 {}

\newcommand{\DLTL}{\emph{DLTL}}

\begin{document}
\pagestyle{empty} %Camera ready without page numbers
\title{Verification of Security Properties of Cryptographic Protocols Through Model Checking Execution Traces}
\titlerunning{Verification of Security Properties of Cryptographic Protocols}
%Anonymized
\author{Authors} 
\author{Simona Bernardi\inst{1}\orcidID{0000-0002-2605-6243} and \\
Joaqu\'{\i}n Ezpeleta\inst{1}\orcidID{0000-0002-9622-8186}
%\and Third Author\inst{3}\orcidID{2222--3333-4444-5555} 
}
%Anonymized
\authorrunning{S. Bernardi and J. Ezpeleta}
\institute{Universidad de Zaragoza, Spain \\
\email{\{simonab,ezpeleta\}@unizar.es}}
\maketitle              % typeset the header of the contribution

\let\thefootnote\relax\footnotetext{
    \textbf{Preprint notice:} 
This is an author-accepted manuscript of a paper accepted for publication in the proceedings of EICC 2026. 
\textcopyright{} Springer. 
}

\begin{abstract}
%The abstract should briefly summarize the contents of the paper in 150--250 words.
Developing cryptographic protocols that are free from vulnerabilities is very challenging, 
as evidenced by numerous design flaws and implementation bugs in widely used protocols. 
%Formal methods have proven effective in uncovering design-level vulnerabilities exploitable by Dolev-Yao attackers
Recent work has leveraged model checking to generate \emph{negative} test cases, that is test cases that include the behavior of Dolev-Yao attacker.
In this paper, we address the problem of analyzing execution traces produced by testing protocol implementations with such negative test cases.
We propose a method for detecting violations of security properties 
%, concretely secrecy and authentication, 
by model-checking execution traces. 
Our approach employs a model checker based on DLTL, a Linear Temporal Logic with freeze operators, and supports automation of the testing process through pre- and post-processing steps. 
These steps, respectively, bridge the gap between different levels of abstraction 
%---namely, the specification of security properties and the execution traces--- 
and synthesize the results produced by the model checker.
We present experimental results as a proof of concept for the feasibility of the proposed method. 

\keywords{Cryptographic protocols  \and Security properties \and DLTL \and Testing}
\end{abstract}
\section{Introduction}\label{s:intro}
%Motivation
Developing cryptographic protocols free from vulnerabilities is very challenging. 
Many protocols %---including widely used ones like TLS, WPA2, Kerberos and OAuth 2.0---
have been repeatedly found to contain design flaws and implementation bugs~\cite{nvd}.
%, as reported by reference standards~\cite{rfc6749,rfc8446}  and public databases of known vulnerabilities~\cite{cve,nvd}.
%Early stage verification of cryptographic protocols with formal methods:   identify vulnerabilities that can be exploited by DY attacks 
Symbolic techniques %, like model checking, 
have been successfully applied primarily to the verification of cryptographic protocols in their early stages  
enabling the identification of vulnerabilities exploitable by Dolev-Yao attacks~\cite{BaBaBhBlCrLiPa_21}, where
%The Dolev-Yao~\cite{DoYa_83} model is the threat model considered by such verification techniques: 
%it assumes that
the attacker has full control over the network.
% ---i.e., he can intercept and retransmit messages on public channels 
%(as a passive attacker), or block, replay, modify, and even generate new messages (as an active attacker).
%The vulnerabilities that can be detected by such techniques are thus flaws in the protocol logic.
The main drawback of tools supporting symbolic analysis is their limited scalability~\cite{BaBaBhBlCrLiPa_21}, which can make the verification of real-world case studies extremely time-consuming. 
%, with analyses potentially taking days depending on the protocol.

%Comment about the weaknesses of these approaches: explosion of the state space, time consuming, undecidability or bounded guarantees.

%Model checking  in testing of protocols under development
At later stages of development, recent works have leveraged model checking to generate test cases for protocol implementations~\cite{GaMo_24,YaXiYiShWaZhJi_25}. 
In particular, in~\cite{GaMo_24} a method is proposed for generating \emph{negative} test cases, i.e., test cases that incorporate the Dolev-Yao intruder. 
A negative test case %---comprising both the specification and the implementation--- 
is automatically derived from counterexamples %, i.e., symbolic execution traces, 
generated by model checking protocol specifications against security goals, such as secrecy and authentication.

%Our contribution
In this paper, we build on~\cite{GaMo_24} and take a step further by addressing the problem of analyzing execution 
traces produced by testing protocol implementations with such negative test cases.
Specifically, we propose a method to determine whether the implementation pass or not a test case by verifying the execution traces against 
a security goal defined in the original protocol specification. 
%Similarly to symbolic-based approaches, the proposed method aims to detect vulnerabilities in protocol logic; 
%however, unlike those approaches, it relies on the verification of execution traces obtained from prototype protocol implementations. 
This technique can serve as an alternative when symbolic techniques  are impractical, for instance due to time constraints.

%DLTL model checking
The analysis is conducted using a model checker based on DLTL logic~\cite{CoEz_19}, an extension of LTL that allows the specification and evaluation of properties involving relations between attribute values across different states. 
Given a protocol execution trace and a DLTL formula encoding the violation of a security goal, the model checker verifies whether the formula is satisfied by the trace.
%,i.e., the security goal is violated. 
%For instance, the DLTL formula encoding a secrecy violation expresses the situation in which an unauthorized agent gains access to a secret.

%Pre/Post-processing 
The proposed method facilitates automation of the testing process by including pre/post-processing steps.
Pre-processing of execution traces establishes the mapping between the bitstrings in exchanged messages 
and the assets in the security goals.
% (specified at the design-level).
Besides, post-processing of the model checking results enables the test outcome to be interpreted in terms of the security goal and, 
when the goal is violated, allows for the identification of the specific traces that breach the property.
%Proof-of-concept: "tool" and the experiments
As a proof of concept, we have developed a tool that has been used in the experiments with a set of protocols from the 
AnBx library~\cite{AnBxweb}. %of case studies 

%Organization of the paper
%\SB{The paper organization updated.}

The paper is organized as follows:
the rest of this section %Section~\ref{s:rw} 
reviews briefly the related work;
Section~\ref{s:overview} gives an overview of the approach;
Section~\ref{s:dltl} defines the DLTL %trace
model and %the DLTL 
formulas for checking security properties;
Section~\ref{s:testing} describes %in details the steps of the
trace checking workflow supported by the \emph{goal\_checker} tool;
Section~\ref{s:experiments} reports the results from the experiments;  and % carried out to validate the tool; and
Section~\ref{s:conclusions} provides conclusions and future work.
%Appendices~\ref{app:anb} and~\ref{app:dltl} provide, respectively, the background on \emph{Alice\&Bob} languages and DLTL.

\subsection{Related Work}\label{s:rw}

%Tools (and techniques) supporting the formal verification of cryptographic protocols via symbolic analysis 
Many tools are available to support the formal verification of cryptographic protocol specifications through symbolic analysis~\cite{BaBaBhBlCrLiPa_21}. 
Given a protocol and a security property to be verified %---such as secrecy or authentication---
these tools typically return an attack trace as a counterexample if the property is violated. 
In this work, we use the %On-the-Fly Model Checker 
OFMC~\cite{BaModVi_05} model checker in the experiments. 
% to validate our results. 
One significant limitation we faced while using the OFMC % in our experiments 
is the verification time, which increases exponentially with the number of protocol sessions.

%Frameworks
Besides, various frameworks have been developed to assist non-experts in formal methods
by providing user-friendly languages for specifying protocols and security properties, while also
integrating multiple analysis tools as backends~\cite{AVISPA_05,AnBxIDE_24}.
In particular, AnBx~\cite{AnBxIDE_24} %is an Eclipse-based framework that 
supports the specification of protocols and properties in  ``Alice\&Bob'' formal languages~\cite{BuCaMoMo_16},
and includes OFMC as a backend. 
Additionally, the AnBx framework provides model-to-code generation features~\cite{Mo_16} for rapid Java prototyping.
%;specifically, a Java-based prototype can be automatically derived from an ``Alice\&Bob'' protocol specification. 
This feature is leveraged in the approach proposed in this work to generate negative test cases.

%Testing / runtime verification
Formal methods have been mainly used in the verification of cryptographic protocol specifications at design stage.
Recent proposals leverage them for fuzzing or runtime verification~\cite{AmHiKr_24,MoKu_24}.
Both works make use of formal logics, similar to the linear temporal logic used in the Tamarin tool~\cite{BaCrDrSa_17}, to express security properties: in~\cite{AmHiKr_24}, the logic is used to guide the fuzzer in identifying violations of security properties caused by Dolev-Yao attacks; 
in~\cite{MoKu_24}, it is used to monitor protocol execution and detect potential violations of security properties.
Although our work differs in focus from~\cite{AmHiKr_24,MoKu_24}, %---as we neither propose a fuzzer nor consider runtime monitoring---
it shares a common challenge: coping with different levels of abstraction, particularly between the specification of security properties and the observation of concrete execution traces.

Linear Temporal Logic (LTL) is a formalism used to specify and reason about the behavior of systems over time, where properties are expressed along a single \emph{trace} (a linear sequence of states or events). It introduces temporal operators such as X (neXt), F (Eventually), G (Globally), and U (Until) to describe how propositions evolve across future points in the trace~\cite{MaPn92}. 

Alur and Henzinger introduced Timed Propositional Temporal Logic (TPTL)~\cite{AlHe1994}, which augments LTL with freeze quantifiers, a mechanism that allows one to bind a variable to a state and subsequently compare these frozen state values with later states, within a temporal formula. This makes TPTL capable of expressing data-dependent temporal constraints, whereas standard LTL only reasons about Boolean propositions without memory of specific values. Thus, TPTL with freeze operators is strictly more expressive than classical LTL. \DLTL\ logic~\cite{CoEz_19} adopts the same perspective as TPTL, but adapting it to the case of finite traces, and also adding past operators.

\section{Approach Overview}\label{s:overview}
\begin{figure}[htb]
\begin{center}
\includegraphics[width=\textwidth]{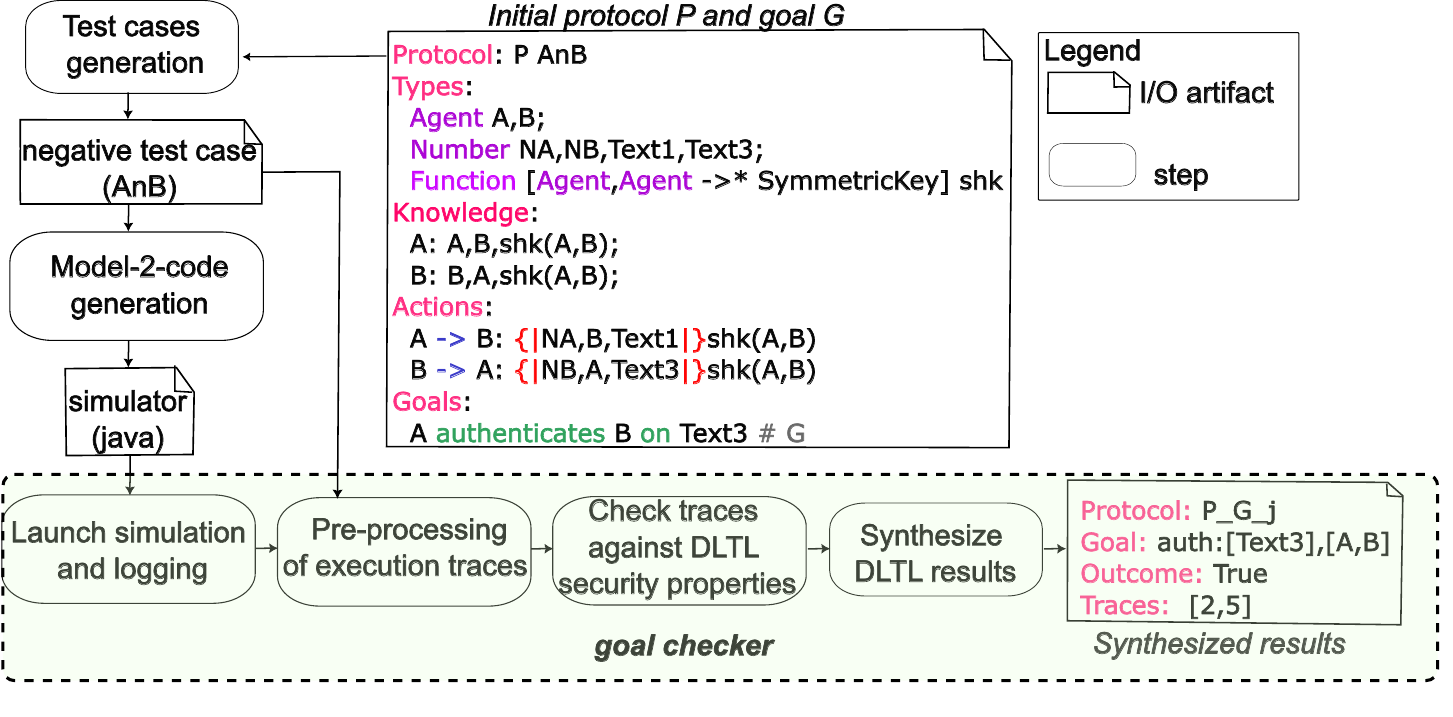}
\caption{Workflow and dataflow of the overall approach}
\label{fig:overview}
\end{center}
\end{figure}

Figure~\ref{fig:overview} shows the overall approach for testing security properties of cryptographic protocols 
with checking of execution traces. 
Our approach relies upon the AnBx framework~\cite{Mo_16} and, in particular, 
the \emph{Alice\&Bob} protocol narration as input specification (AnB) and the automatic generation of a protocol Java-based implementation.
%Appendix~\ref{app:anb} provides background on the \emph{Alice\&Bob} languages.

Given an initial AnB specification of a protocol $P$ and a goal to be verified $G$, a set of negative test cases is considered.
Each test case  $P\_G{_j}$ is an AnB specification that, besides the original actors and goal, includes an intruder who may invalidate the property of interest. 
A test case can be semi-automatically generated (\textit{Test cases generation}): 
indeed, the AnBx framework already provides support to automatically generate test cases~\cite{GaMo_24}
albeit only for a subset of attack scenarios, due to the limited expressivity of the \emph{Alice\&Bob} languages.
%This limitation is illustrated by an example at the end of this section.
Similarly, a Java-based implementation of a test case can be semi-automatically generated with the AnBx
framework (\textit{Model-2-code generation}).

%Contribution of this paper
The \emph{goal checker} workflow, emphasized by the dotted rounded rectangle, addresses the 
verification of the goal $G$ through DLTL checking of the protocol execution traces and it is the focus of this paper.

The Java-based implementation of a negative test case $P\_G{_j}$ is used to launch a protocol simulation and log the execution traces: 
a trace represents the execution of a session of the protocol. 
Then, the execution traces are pre-processed to get a structured log  and the goal as DLTL formula (i.e., goal violation),
which are fed to the DLTL model checker. 
The latter produces outcomes %(i.e., true/false) 
for each trace in the log against the DLTL security property: the results are finally
synthesized to summarize the outcome of the verification of the log against of the original goal. A list of traces in the log that violated the
goal are listed in the summary.

\section{Specification and Verification with DLTL}
\label{s:dltl}

To verify that protocol executions comply with their intended specifications, we employ DLTL~\cite{CoEz_19}. In our context, a trace is defined as a chronologically ordered sequence of events representing a protocol's execution. Each event encapsulates the metadata of a communication action, such as the participating agents, the action type, and the associated parameters, represented as typed attributes.
%(e.g., integers, strings, sets).

As we will demonstrate, protocol requirements often go beyond than simple sequences of actions, requiring the correlation of attribute values across different states in the trace. For instance, a property might specify that a nonce identifier captured during an initial event must be identical to the identifier appearing in a subsequent event.

By utilizing freeze operators, we can evaluate these cross-temporal properties without prior knowledge of the data domain (e.g., the set of all possible nonce IDs). While traditional LTL would require a distinct proposition for every possible value, leading to an unmanageable or even infinite state space, freeze operators allow us to bind an event to a variable and perform direct comparisons later in the trace. This makes this kind of logic to be well-suited for verifying protocols where the specific data values are not known in advance, but the consistency of those values is critical.

%This is the reason for adopting a logic with freeze operators, such as DLTL. 
%A formal and complete definition of the syntax and semantics of this logic can be found in Appendix~\ref{app:dltl}.

\subsection{DLTL Model}\label{ss:dltl_model_formula}
%In the second sub-step, carried out by the \emph{dltl\_generator} module, both a DLTL \emph{trace model} and 
%a DLTL parametric security property are generated.
The DLTL %trace 
model $\sigma = e_1e_2\dots e_n$ is the ordered sequence of events $e_i$, where each $e_i$ corresponds to the $i^{th}$ event of the log
and is a tuple of the form:
$$e_i \equiv \{ a^{action}_i,n^{session}_i,n^{step}_i,s^{active}_i,s^{passive}_i,@^{mess}_i,s^{type}_i \}.$$
Each element of the tuple corresponds to an attribute %of the $i^{th}$ event 
as shown in Table~\ref{tab:event_attributes}:
in particular, $a^{action} \in \mathcal{A}$ is the action ($\mathcal{A}$ is the set of actions, described in Table~\ref{tab:actions}), % (atomic proposition), 
$n^{session}$ and $n^{step}$ are the session and step numbers respectively, %(integers),
$s^{active}$ and $s^{passive}$ are the active and passive participants, %(string),
$@^{mess}$ is the message content, and  %(set of strings), and
$s^{type}$ is the type of the message. %(string).

\begin{table}
%{\scriptsize
\begin{tabular}{|p{2cm}|p{7cm}|p{3cm}|}
	\toprule
	\textbf{Attribute} & \textbf{Description} & \textbf{DLTL Domain}\\\midrule
%	timestamp & Time stamp of the logged event (zero values are missing timestamps) 
%\SB{we  omit timestamps, since they are not used in this work}\\\midrule
	session & Number of the session & integer\\\midrule
	step & Number of the step & integer\\\midrule
	active & Concrete agent who  performs the action & string\\\midrule
	passive & Concrete agent who is involved in the action  & string\\\midrule
	action & Type of action (cfr. Table~\ref{tab:actions}) & atomic proposition\\\midrule
	mess & Content of the message & set of strings\\\midrule
	type & The type of the message  & string\\
	\bottomrule
\end{tabular}
%} %scriptsize
\caption{Event attributes.}
\label{tab:event_attributes}
\end{table}

The possible actions are listed in Table~\ref{tab:actions}. In addition to the send and receive actions, the remaining ones are related to message preparation on the sender side or post-processing on the receiver side. 
%Although their names are specific to the prototype's runtime logging output format, they can be easily adapted to other implementations.

\begin{table}[htb]
%{\scriptsize
\begin{tabular}{|p{3cm}|p{9,5cm}|}
\toprule
\textbf{Action} & \textbf{Description}\\\midrule
sent & A message is sent \\ %by the active participant to the passive participant\\
received & A message is received \\ %by the active participant from the passive participant\\
generateNumber$^*$ & The active participant generates a fresh value  \\
AnBx\_Params & The active participant prepares the message content to be sent\\
eq\_check$^*$ & The active participant verifies the message content with an equality check\\
inv\_check$^*$ & The active participant verifies the message content with a crypto-operation (e.g., decryption) or a projection operation followed by another check
(e.g., extracts an element of the tuple and performs an equality check)\\\midrule
\multicolumn{2}{l}{
{\scriptsize $^*$It can be postfixed with the name of an \emph{asset} (Agent or Number type) in the AnB specification.}
}\\
\bottomrule
\end{tabular}
%} %scriptsize
\caption{Type of actions in the events of a trace.}
\label{tab:actions}
\end{table}

\subsection{DLTL Formulas}
%The  \emph{dltl\_generator} module also generates the DLTL  security property that will be used by the model checker to verify
%the goal specified in the AnBx test case protocol. 
In this paper, we use DLTL formulas to represent goal violations. Depending on the goal a different DLTL formula is set as detailed in the following. 

\paragraph{Authentication.}
Let us consider the goal \texttt{"A authenticates B on N"}, where \texttt{A} and \texttt{B} are two agents and \texttt{N} is an asset
%\footnote{We name as \emph{asset} a variable of type \emph{Number} in the AnB specification of the protocol.}  
exchanged in the protocol. %---such as nonce \texttt{Text3} in  Listing 1.1.
The authentication goal is violated when \texttt{B} does not agree on \texttt{N} or eventually \texttt{A} is not able to recover \texttt{N}.
The DLTL formula~(\ref{eq:auth}) expresses such a violation:
%;GOAL auth:[?NB],[?A,?B]
%G z.( !("(z)z[active]==?B" & ?agreesNB) |! X F y.("(y)y[active]==?A" & ?can_recover & ("(z,y)z[mess]==y[mess]")))
%\begin{framed}
%{\small
\begin{eqnarray}\label{eq:auth}
\phi_{A,B,N} & \equiv \mathrm{G\ z}.\Big( &  \neg agrees(\mathrm{z})\ \vee  
    \neg\ \mathrm{X F\ y}.\big(   can\_recover(\mathrm{z},\mathrm{y})\big)\Big)
\end{eqnarray}
%}
%\end{framed}
where $agrees(z)$ and $can\_recover(z,y)$  are %conjunctive compositions of 
boolean functions: the former returns
$true$ if and only if there is an event where $B$  can compose $N$, the second returns 
$true$ if and only if there is an event where $A$ can recover $N$ and  the agreed and recovered values are equal. Formally:
\begin{eqnarray}
agrees(\mathrm{z}) & \equiv &  \big(  s^{active}_z == B\ \wedge\ can\_composeN \big)\\
recover(\mathrm{y}) & \equiv & \big( s^{active}_y == A\ \wedge\ can\_recover\ \wedge\ @^{mess}_z == @^{mess}_y \big) 
\end{eqnarray}
The propositions $can\_composeN$ and $can\_recover$ are defined in Table~\ref{tab:component}.
In particular, $can\_composeN$ is defined as the disjunction of three actions listed in Table~\ref{tab:actions}, each postfixed with the name of the asset included in the goal, i.e., \texttt{N}. Depending on the protocol specification, one of these actions corresponds to the event in which \texttt{B} either generates or recovers the asset.

In case of a goal including several assets, i.e., \texttt{"A authenticates B on N\_1,...,N\_n"}, the DLTL formula is generalized as follows:
\begin{equation}
\phi_{A,B,\{N_1,\dots,N_n\}} \equiv \mathop{\vee}_{\alpha=1}^n \phi_{A,B,N_\alpha}
\end{equation}

\begin{table}[htb]
\begin{center}
\begin{tabular}{p{3cm}|p{9.5cm}}
\toprule
Function & Definition\\\midrule
$can\_composeN$ & $a^{action} == (generateNumberN \vee inv\_checkN \vee eq\_checkN)$\\
$can\_recover$ & $a^{action} == (inv\_check\ \vee\ eq\_check)$\\
$generatesN$ & $a^{action} == generateNumberN$\\
%$in\_mess(N)$ &  $N \in @^{mess}$ \\
$can\_decrypt$ &  $a^{action} == inv\_check\ \wedge$\\
			& $\neg (s^{type} == ``javax.crypto.SealedObject.toString")$\\
\bottomrule
\end{tabular}
\end{center}
\caption{Component boolean functions}
\label{tab:component}
\end{table}

\paragraph{Secrecy.}
Let us consider the goal \texttt{"N secret between A\_1,\dots,A\_m"}, where \texttt{A\_1,\dots,A\_m} is a set of $m$ agents and \texttt{N} is an asset exchanged in the protocol.
The secrecy goal is violated if one agent of the set generates \texttt{N} and someone different from 
the agents in the set can read \texttt{N}. 
The DLTL formula~(\ref{eq:secrecy}) expresses such a violation:
%;SECRET_VIOLATION
%F z.(("(z)z[active]==?A" | "(z)z[active]==?B") & ?generatesNxNA & X F y.(("(y)y[active]!=?A") & ("(y)y[active]!=?B")& ?can_read & ("(z,y)z[mess]==y[mess]") & ("(y)y[type]!='javax.crypto.SealedObject.toString'")))
%\begin{framed}
%{\small
\begin{equation}\label{eq:secrecy}
\psi_{N,\{A_1,\dots,A_m\}} \equiv \mathrm{F\ z}.\Big( generates(\mathrm{z})\ \wedge\ \mathrm{X\ F\ y}.\big(  leaked(\mathrm{z},\mathrm{y}) \big) \Big) 
\end{equation}
%}
%\end{framed}
where $generates(\mathrm{z})$ and  $leaked(\mathrm{z,y})$ are %conjunctive compositions of 
boolean functions: 
the former returns $true$ if and only if there is an event where an agent from the set $\{ A_\alpha\}_{\alpha=1,\dots,m}$ is generating $N$, and the latter
returns $true$ if and only if there is an event where an agent $R$, with $R \not\in \{A_\alpha\}_{\alpha=1,\dots,m}$,
can successfully decrypt $N$. Formally:
\begin{eqnarray}\label{eq:generates}
generates(\mathrm{z}) 	&\equiv 	& \Big( \mathop{\vee}_{\alpha=1}^m s^{active}_z == A_\alpha\ \wedge generatesN   \Big)\\\label{eq:leaked}
leaked(\mathrm{z},\mathrm{y}) 		& \equiv & \Big( \mathop{\wedge}_{\alpha=1}^m \neg (s^{active}_y == A_\alpha) \wedge can\_decrypt \wedge @^{mess}_z == @^{mess}_y  \Big)\\\nonumber			
\end{eqnarray}
The propositions $generatesN$ and $can\_decrypt$ are defined in Table~\ref{tab:component}. The former denotes the nonce-generation action listed in Table~\ref{tab:actions}, postfixed with the name of the asset included in the goal, i.e., \texttt{N}. The latter identifies the event in which a decryption operation is successfully performed and the resulting message is not sealed.

\section{The Goal Checker Workflow}\label{s:testing}

In this section, we describe in detail the steps of the \emph{goal checker} workflow of Figure~\ref{fig:overview} together with the
Python modules of the tool developed as a support.
% is refined in
%Figure~\ref{fig:goal_checker}: each swim-line of the diagram corresponds to a Python module of the tool\footnote{URL: removed to comply with the anonymity submission policy.} developed as a support of the workflow.

%\begin{figure}[htb]
%\includegraphics[width=\textwidth]{figs/goal_checker.pdf}
%\caption{Goal checker workflow (refined).}
%\label{fig:goal_checker}
%\end{figure}

\subsection{Launch Simulation and Logging}
The first step of the workflow aims to produce the execution traces of a negative test case protocol.
Our approach relies upon the AnBx framework~\cite{Mo_16} to generate a Java-based implementation of the protocol.
The \emph{sim\_launcher} module is in charge of compiling the protocol and %considering the AnBx dependencies, 
launching it, by running a concurrent process  for each protocol agent (i.e., role)
and saving the execution traces of each process in a separate textual file.
For each role $R$ in the protocol, as many execution traces $\{ t^{R}_i \}_{i=1,\dots,n}$ are generated as the number of sessions $n$ 
specified as configuration parameter: each trace  $t^{R}_i$ includes a set of ordered events involving the concrete agent  
playing the role $R$ in the protocol (i.e., instantiated value of the role $R$). 
%For example, considering the negative test case \texttt{P\_G\_1} (Listing 1.1) and assuming a simulation of $10$ sessions
%of the protocol, then three sets of execution traces are generated, one for each role:  $\{ t^{A}_i \}_{i=1,\dots,n}, \{ t^{I}_i \}_{i=1,\dots,n}, \{ t^{B}_i \}_{i=1,\dots,n}$, each set includes $n=10$ traces.

\subsection{Pre-processing of Execution Traces}\label{ss:preprocessing}
The execution traces produced by the protocol need to be pre-processed to be fed to the DLTL model checker.
This step produces the formatted traces and the DLTL formulas in two sub-steps as follows.
%In the following, the two sub-steps
% which are carried out sequentially by two different modules are described.

\subsubsection{Extract Log and Map Agents to Protocol Roles.}
In the first sub-step, carried out by the \emph{log\_extractor} module, 
the execution traces of each role are filtered, considering the type of actions listed in Table~\ref{tab:actions}.
The filtered execution traces of the roles are then merged together to get a unique log, where the events are ordered according to the session number 
(i.e., trace identifier) and the number of the step. In particular, the latter corresponds to the number of the action in the AnB specification of the protocol.
%:for example, in the negative test case \texttt{P\_G\_1} (Listing 1.1) there are four actions numbered from $0$ to $3$.
Then, the events in the log are formatted and unique identifiers are assigned to asset values %---prefixed by \texttt{X}--- 
as well as to encrypted messages. %---prefixed by \texttt{@}.

%\newpage
\begin{lstlisting}[caption={Excerpt of the log produced by the \emph{log\_extractor}},label={list:log},numbers=left]
#session,step,active,passive,action,mess,type
1,0,charlie,-,generateNumber,X0,anbxj.Crypto_ByteArray
1,0,charlie,-,generateNumber,X1,anbxj.Crypto_ByteArray
1,0,charlie,-,AnBx_Params,X0;bob;X1,anbxj.AnBx_Params
1,0,charlie,alice,sent,@3b6eb2ec,javax.crypto.SealedObject
1,0,alice,charlie,received,@3b6eb2ec,javax.crypto.SealedObject
1,0,alice,-,inv_check,@c2e1f26,javax.crypto.SealedObject.toString
\end{lstlisting}
The Listing~\ref{list:log} shows an excerpt of the log produced by the \emph{log\_extractor}  from the execution traces of a negative test case.
In particular, the logged events correspond to the exchange of a message encrypted with the secret key shared by roles $A$ and $B$.
The message is a triplet consisting of a nonce $NA$, the role name $B$ and  nonce $Text1$.

%Beside the header (line 1), there are six events in the log related to this action:
The first three events (lines 2-4) are actions performed by \texttt{charlie} (role $A$)  to prepare the message, i.e., generating the two nonce values \texttt{X0} and \texttt{X1}, respectively, and setting the triplet \texttt{\{X0,bob,X1\}}, where \texttt{bob} plays the role $B$. 
The lines 5-6 correspond to the  sending and the reception of the encrypted message  \texttt{@3b6eb2ec},
%---as the message AnBx type indicates (i.e, it is a sealed object).
where the receiver is \texttt{alice} ---who plays the role of the intruder $I$. 
The last event (line 7) is the action performed by \texttt{alice}, who tries to decrypt the message without success
(the \texttt{inv\_check} operation returns an encrypted value, i.e., \texttt{@c2e1f26}).

The \emph{log-extractor} module is also responsible for mapping the agents in the log to the roles defined in the AnB protocol and 
for generating the component boolean functions listed in Table~\ref{tab:component}, taking into account the goal to be verified.
%This mapping will be used in the following steps to generate the DLTL security properties to be checked.
%Listing~\ref{list:macro1} shows the macro definitions of the roles (lines 1-3) and the boolean functions (lines 4-5) 
%for the negative test case protocol \texttt{P\_G\_1}.

%\begin{lstlisting}[caption={Macro definitions and  DLTL security property (\texttt{P\_G\_1} protocol)},label={list:macro1},numbers=left]
%_SET ?A 'charlie'
%_SET ?B 'bob'
%_SET ?I 'alice'
%_SET ?can_composeText3 (generateNumberText3 | inv_checkText3 | eq_checkText3)
%_SET ?can_recover (inv_check | eq_check)

%;GOAL auth:[?Text3],[?A,?B] 
%G z.( !("(z)z[active]==?B" & ?can_composeText3) |
%      ! X F y.( "(y)y[active]==?A" & ?can_recover & ("(z,y)z[mess]==y[mess]")
%              )
%    )
%\end{lstlisting}

\subsubsection{Generate DLTL Model and Formula.}
In the second sub-step, the \emph{dltl\_generator} module generates both the DLTL model and the DLTL formula.
The DLTL model closely resembles the log produced in the first step %(see Listing~\ref{list:log}) 
but conforms to the specific input format required by the DLTL tool. 
The DLTL formula is generated based on the goal specified in the negative test case.
%For example, the Listing~\ref{list:macro1} (lines 8-11) shows the DLTL formula (\ref{eq:auth}), in the DLTL tool specific syntax, 
%generated from the authentication goal specified in \texttt{P\_G\_1} (Listing 1.1) and reported as comment in line 7.

\subsection{Check Traces and Synthesize Results}

%The boolean functions defined in the DLTL formulas~\ref{eq:auth} and~\ref{eq:secrecy} are parametric with respect to the protocol roles in the 
%goal ---i.e., $A$, $B$. 
The \emph{dltl\_model\_checker} is in charge of checking the DLTL formula against each trace in the DLTL model.
%takes as input the model and the formula generated by the \emph{dltl\_generator} module: 
%it replaces the \texttt{\_SET} macros with their specifications in the security property and then checks the formula against each trace in the model.
As a result, it produces a boolean value for each trace. % in the log.
In the last step, performed by the \emph{res\_synthesis} module, the results are synthesized in order to have a unique boolean outcome for all the traces in model.
In particular, if at least a trace $i$ satisfies the formula then the outcome is $true$ and the trace identifier $i$ is 
added to the list of traces violating the goal.
Figure~\ref{fig:overview} shows the synthesized result produced for the negative test case  $P6\_A2$ in Table~\ref{tab:protocols},
that represents a replay attack, considering a log with ten traces, 
where two out of ten traces violate the authentication goal.

\section{Experiments}\label{s:experiments}

As a proof of concept for the feasibility of the proposed method, we have developed a tool\footnote{Available at: https://github.com/simber72/goal-checker/} that has been validated with a set of protocols from the AnBx library~\cite{AnBxweb}.
% Original protocols and goals
Table~\ref{tab:protocols}, shows the protocols $P_i$ used in the experiments with their original name %, for reference purpose, 
(column 1) and the goals (column 2). 
%For some protocols, more than one goal is considered. ---i.e., $P_5,P_6$ and $P_{8}$.
%Violated column
The AnB specification of each protocol has been verified against the goals with OFMC~\cite{BaModVi_05}, as a result
%(column 3) 
all the protocols are vulnerable to attacks against authentication or secrecy. %, but the protocol $P_2$.

%Test cases with MITM
For each protocol and goal, we have considered at least a negative test case (column 3) involving an intruder whose behavior is intended to violate the corresponding goal and follows an attack pattern: 
\emph{passive}, reading the message in transit;
\emph{replay}, forwarding a received message %containing the same nonce 
multiple times across different protocol sessions;
\emph{reflection} receiving a message from a participant and sending it back;
\emph{mitm}, impersonating one or more legitimate parties in the protocol execution.
%Test cases automation: detailed comments in the conclusions as future work
Most of the test cases were generated in an semi-automatic manner with the support of the AnBx tool~\cite{GaMo_24}.
%, which provides such functionalities given the counter-example of a security goal generated by the OFMC model checker. 
%Rationale on test cases and n.traces
Depending on the type of goal, either a single session ---for a secrecy goal--- or multiple sessions ---for an authentication goal--- were simulated (column 4). 

\begin{table}[htb]
\begin{center}
{\scriptsize
\begin{tabular}{|p{4cm}|c|c|c|c|}
\toprule
Protocol ($P$) 	& Goal$^*$ 	 	& Test case  	& Sessions 	& Results \\
			& 		& 	 (attack type) 	&	&  (goal-violating traces) \\
\midrule
AndrewSecureRPC ($P_1$)     & $A$		& $P_1\_A$ (replay) & 10 & \cmark ($2,3,4,5,8,10$) \\ %'02', '03', '04', '05', '08', '10'
From\_A\_and\_Back ($P_2$)    & $S$ 		& $P_2\_S$ (passive) & 1 & \cmark  \\
GSM ($P_3$) 				& $S$ 		& $P_3\_S_1$ (passive) & 1 & \cmark\\
						& 	     	        & $P_3\_S_2$ (mitm) & 1 & \cmark \\
NSPK ($P_4$) 				& $S$		& $P_4\_S_1$ (passive) & 1 & \xmark \\
						& 	    		& $P_4\_S_2$ (mitm) & 1 & \cmark \\
%WooLam1 ($P_5$) 			& $A$ 		& $P_5\_A_1$ (passive) & 10 & \xmark\\
%						& 	    		& $P_5\_A_2$ (mitm) & 10 & \cmark (all) \\
ISOPK2PMAP ($P_5$) 	& $A1$  	& $P_5\_A1$ (reflection) & 10 & \cmark (all)\\
							 		& $A2$  	& $P_5\_A2$ (reflection) & 10 & \cmark ($2,4,6,8,10$)\\ %['02', '04', '06', '08', '10']
ISOSK2PMAP ($P_6$) 	& $A1$  	& $P_6\_A1$ (replay) & 10 & \cmark ($2,4,9,10$)\\ %['02', '04', '09', '10']
									& $A2$  	& $P_6\_A2$ (replay) & 10 & \cmark ($2,5$)\\ %['02', '05']
Kerberos ($P_7$) 						& $A$  	& $P_7\_A$ (mitm) & 10 & \cmark (all)\\
%KeyEx4b ($P_9$) 						& $A$  	& $P_9\_A$ (mitm) & 10 & \cmark ($2,3,4,6,8,10$)\\
SSO ($P_{8}$) 							& $A1$  	& $P_{8}\_A1$ (mitm)& 10 & \cmark (all)\\
									& $A2$  	& $P_{8}\_A2$ (replay) & 10 & \cmark ($3,5$)\\ %['03', '05']
									& $S$	& $P_{8}\_S$ (mitm) & 1 & \cmark\\
\midrule
\multicolumn{4}{l}{
{\scriptsize * $S$ is a secrecy goal and $A,A1$ and $A2$ are authentication goals.}
}\\
\bottomrule
\end{tabular}
}
\end{center}
\caption{Results from the experiments}
\label{tab:protocols}
\end{table}

%Comments on the results
The last column shows the results, % obtained from the \emph{goal\_checker} tool.
which are consistent with those produced by OFMC. 
Indeed, for all protocols, the \emph{goal\_checker} finds traces of at least one negative test case that satisfy the DLTL formula associated with
the goal---that is, violate the goal.
%This holds true also for protocol $P_2$, specifically for the trace associated with test case $P_2\_S_2$.
%It is worth noting that the test case $P_2\_S_2$ was manually made and represents a scenario in which the legitimate party initiates the protocol with the intruder mistakenly believing him to be a legitimate participant.
Negative test cases aimed at checking authentication goals have been run for several sessions, i.e., $10$ sessions, to assess strong authentication: while
checking strong authentication with OFMC is quite time-consuming ---even for very few sessions, such as 2 or 3, the time to check is in the order of minutes or greater--- the goal\_checker takes few seconds to analyze $10$ traces\footnote{Experiments have been conducted using a MacOS laptop with 
2,3 GHz Intel Core i7 with 4 kernels.}.

\section{Conclusions}\label{s:conclusions}
We have presented a method for testing the security properties of cryptographic protocols by model-checking execution traces. 
The model-checker takes an execution trace and a DLTL formula expressing the violation of a security goal %---i.e., secrecy or authentication---  
and verifies whether the formula is satisfied by the trace.
Pre-processing of execution traces and post-processing of the results facilitate the automation of the testing process.

The experiments were conducted using simple protocols from the AnBx library;
as future work, we are planning to apply the method to a real-world case study in the critical infrastructure domain.

%The proposed approach is implementation independent, provided that the implementation can log the event attributes required to check the DLTL formulas (i.e., the agent performing the action, the action performed, the message exchanged, and the message type). In particular, the modular design of the tool facilitates reuse in different implementation settings, where only the actions and message types are implementation specific and can be easily adapted within the log-extractor module.

%Future work: to be polished
Achieving full automation in the generation of negative test cases remains an open challenge, which we are currently working to address.
%For the specification and Java-based implementation of test cases, we relied on the AnBx framework. 
%This choice allowed us to concentrate on analyzing execution traces using the DLTL model checker and to develop a supporting tool, 
%which we validated through our experiments.
We recognize limitations in the expressiveness of \emph{Alice\&Bob} languages which prevents the specification of negative test cases with
attacks that involve multiple protocol sessions (such as replay and reflection attacks), and we aim to overcome them.

%Anonymized
\begin{credits}
\subsubsection{\ackname} 
The authors were supported by the Spanish Ministry of Science and Innovation [ref. PID2020-113969RB-I00, PID2024-159323NB-I00] and
by the European Union-Spanish ``Agencia Estatal de Investigaci\'on" [ref. PID2023-148202OB-C22].
%PID2023-148202OB-C22, European Union-Spanish 
%Detection and mitigation of vulnerabilities in critical systems (Detección y mitigación de vulnerabilidades en sistemas críticos) - DAMOCLES-PID2020-113969RB-I00. Financed by Ministerio de Ciencia e Innovación
%PID2024-159323NB-I00. UN MARCO FORMAL PARA EL MODELADO, ANALISIS Y OPTIMIZACION DE REDES BIOLOGICAS

\subsubsection{\discintname}
%It is now necessary to declare any competing interests or to specifically
%state that the authors have no competing interests. Please place the
%statement with a bold run-in heading in small font size beneath the
%(optional) acknowledgments\footnote{If EquinOCS, our proceedings submission
%system, is used, then the disclaimer can be provided directly in the system.},
%for example: 
The authors have no competing interests to declare that are relevant to the content of this article. 
% Or: Author A has received research
%grants from Company W. Author B has received a speaker honorarium from
%Company X and owns stock in Company Y. Author C is a member of committee Z.
\end{credits}
%\newpage
 \bibliographystyle{splncs04}
 \bibliography{biblio}
%

%Removed
%\appendix
%\input{sections/07_app}

\end{document}